\documentclass[epj]{svjour}
\usepackage{graphics}
\newcommand{\ltap}{\raisebox{-0.6ex}
                       {$\ \textstyle \stackrel{\textstyle <}{\sim}\ $}
                      }
\newcommand{\gtap}{ \raisebox{-0.6ex}
                       {$\ \textstyle \stackrel{\textstyle >}{\sim}\ $}
                      }
\begin{document}
\title{Indications for Large Rescattering in Rare B Decays}
\author{George W.S. Hou %\inst{1} \and Second author\inst{2}% etc
}                     % Do not remove
\institute{Department of Physics, National Taiwan University,
Taipei, Taiwan 10764 % \and the second here
}
\date{Received: date / Revised version: date}
\abstract{ The sign of $A_{\rm CP}(K^-\pi^+) < 0$, the evidence
for $\bar B^0 \to \pi^0\pi^0$, and the possibly sizable
$A_{\pi\pi}$ and $S_{\pi\pi}$ in $\bar B^0 \to \pi^+\pi^-$ all
suggest that final state rescattering may be needed in $\bar B\to
PP$ decay, which is echoed by large color suppressed $\bar B^0\to
D^0h^0$ modes. An SU(3) formalism of $\mbox{\bf 8}\otimes
\mbox{\bf 8} \to \mbox{\bf 8}\otimes \mbox{\bf 8}$ rescattering in
$PP$ final states leads to interesting predictions, in particular
allowing for small $\bar B^0 \to K^-K^+$.
 % end of PACS codes
} %end of abstract
\maketitle
\section{Motivation}
\label{mot}

Around 1999, the emergence of large $K\pi/\pi\pi$ ratio in $B$
decay lead to the suggestion~\cite{HHY} that maybe $\gamma \equiv
\phi_3 \equiv \arg V_{ub}^* \gtap 90^\circ$, in contrast to the
CKM fit (to other data) of $\sim 60^\circ$. The pattern of $K\pi$,
$\pi\pi$ data can then be understood within factorization. As the
final results from CLEO came out, it was further
speculated~\cite{HY} that rate and (direct) $A_{\rm CP}$ pattern
could hint at rescattering in final state (FSI). If one makes a
final state isospin decomposition, FSI phases are {\it in
principle} present. As $A_{\rm CP}$s depend critically on
absorptive parts, the CP invariant FSI phases could easily shift
direct CP patterns.

The discovery of color suppressed $\bar B^0\to D^0h^0$
decays~\cite{D0h0} above factorization predictions suggests that
FSI may have to be taken seriously.
More recently~\cite{LP03}, the 3$\sigma$ effect of $A_{\rm
CP}(K^-\pi^+) < 0$, the evidence for $\bar B^0 \to \pi^0\pi^0$,
and the possibly sizable $A_{\pi\pi}$ and $S_{\pi\pi}$ in $\bar
B^0 \to \pi^+\pi^-$ etc., all could be hinting at presence of
sizable rescattering in $\bar B\to PP$ final states, where $P$
stands for an octet pseudoscalar. We thereby revisit the FSI
speculation.

Treating the color suppressed $\bar B^0\to D^0h^0$ modes as an
exercise, we developed~\cite{CHY} an SU(3) based $\mbox{\bf
3}\otimes \mbox{\bf 8} \to \mbox{\bf 3}\otimes \mbox{\bf 8}$
rescattering in $DP$ final states. Here we report the
results~\cite{CHY2} on extending the formalism to $PP$ final
states.

\section{\boldmath Ansatz:  Multimode Fit with FSI Phases}
\label{ansatz}

Since factorization seems to account for the rates of leading
decays, we adopt the simple and physical picture of (naively)
factorized amplitudes ${\cal A}^f_{l}$ followed by FSI, i.e.
\begin{equation}
\langle i; \mbox{\rm out}|H_{\rm W}|B\rangle
 = \sum_l {\cal S}^{1/2}_{il} {\cal A}^f_{l}.
\label{eq:master}
\end{equation}
We use naive factorization not just for sake of simplicity, but
because more sophisticated treatment in, say, QCD factorization
introduces hadronic parameters, and one may incur double counting.
Note that $l$ is summed over quasi-elastic channels in Eq. (1). We
assume that the large cancellations between numerous {\it
in}elastic channels generate only the ``perturbative" FSI phase
accounted for by the penguin absorptive part.

We treat $\bar B \to PP$ final states only, since $VP$ modes are
not yet settled (both experiment and theory). Also, we are yet
unable to treat $\eta^\prime$ hence take $\eta \cong \eta_8$.
Thus, we consider $\mbox{\bf 8}\otimes \mbox{\bf 8}\to \mbox{\bf
8}\otimes \mbox{\bf 8}$ rescattering. Since only the $\mbox{\bf
1}$, one of the $\mbox{\bf 8}$s, and the $\mbox{\bf 27}$ are
symmetric, the ${\cal S}^{1/2}$ matrix in Eq.~(\ref{eq:master})
takes up the form
\begin{equation}
{\cal S}^{1/2}=
 e^{i\delta_{\bf 27}} |{\bf 27}\rangle\langle {\bf 27}|
+e^{i\delta_{\bf  8}} |{\bf  8}\rangle\langle {\bf  8}|
+e^{i\delta_{\bf  1}} |{\bf  1}\rangle\langle {\bf  1}|,
\label{eq:Smatrix}
\end{equation}
hence there are just two physical phase differences, which we take
as $\delta \equiv \delta_{\bf 27} - \delta_{\bf 8}$ and $\sigma
\equiv \delta_{\bf 27} - \delta_{\bf 1}$. These FSI phases
redistribute  ${\cal A}^f_{l}$ according to Eq. (\ref{eq:master}).
Alternatively, they can be viewed as {\it a simple two parameter
model} extension beyond the usual $B\to PP$ amplitudes.

It is important to point out that the $\sigma$ phase appears only
in the $\pi^-\pi^+$, $\pi^0\pi^0$, $K^-K^+$, $K^0\overline K^0$,
$\pi^0\eta_8$ and $\eta_8\eta_8$ rescattering subset. It arises
from a total (or double) annihilation of the incoming $PP$ state,
i.e. $(q_1\bar q_2)(\bar q_1q_2) \to (q_1^\prime\bar
q_2^\prime)(\bar
q_1^\prime q_2^\prime)$. %, which we parametrize as $\tilde r_a$.
Heuristically, one has three other type of ``topologies":
``pomeron", with exchange of only energy-momentum; ``exchange",
where a (anti-)quark pair is exchanged; ``annihilation", where a
quark-antiquark pair is annihilated. In the end, besides an
overall magnitude and phase, one is left with two phase
differences.

By adding the two strong phases $\delta$ and $\sigma$, we follow
the multimode fit approach of Ref.~\cite{HSW}. We take only the
better measured or known quantities as input: 7 rates from $K\pi$,
$\pi\pi$, 3 asymmetries from $K^-\pi^+$, $K^-\pi^0$ and $\bar
K^0\pi^+$, and the (theoretical) form factor ratio
$F_0^{BK}/F_0^{B\pi}=(0.9\pm 0.1) \, f_K/f_\pi$.
The fit parameters are $F_0^{BK}$, $1/m_s^{\rm eff}$, $\delta$ and
$\sigma$, and possibly $\phi_3$, where we explore the two cases of
keeping it free (Fit~1), or fixed (Fit~2) at $60^\circ$.
The fit output are the rates, $A_{\rm CP}$s, and especially
$A_{\pi\pi}$ and $S_{\pi\pi}$. These, together with the inputs,
are given in Table~\ref{tab:output}.

\begin{table*}[t!]
\caption{ World average inputs and fitted outputs; data in
brackets are not used in fit, while $\eta_8 K(\pi)^-$ entries are
for $\eta K(\pi)^-$. Horizontal lines separate rescattering
subsets. Fit 1 or 2 stand for $\phi_3$ free or fixed at
$60^\circ$. Setting $\delta = \sigma = 0$ but keeping other
parameters fixed give the results in parentheses; the fitted
parameters and $\chi^2_{\rm min.}$ are given in
Table~\ref{tab:phase}.
 \label{tab:output} }
%\begin{ruledtabular}
\begin{center}
\begin{tabular}{lcrrcrr}
\hline\noalign{\smallskip}
 Modes
 & ${\mathcal B}^{\rm expt}\times10^6$
 & ${\mathcal B}^{\rm Fit1}\times10^6$
 & ${\mathcal B}^{\rm Fit2}\times10^6$
 & $A_{\rm CP}^{\rm expt}~(\%)$
 & $A_{\rm CP}^{\rm Fit1}~(\%)$
 & $A_{\rm CP}^{\rm Fit2}~(\%)$\\
\hline\noalign{\smallskip} $K^-\pi^+$
        & $18.2\pm0.8$
        & $19.4^{+1.0}_{-1.2}$ $(19.7)$
        & $18.5\pm0.6$ $(18.1)$
        & $-9\pm 4$
        & $-6^{+2}_{-3}$ $(9)$
        & $-4\pm 1$ $(7)$\\
$\overline K {}^0\pi^0$
        & $11.2\pm1.4$
        & $8.3^{+1.3}_{-0.4}$ $(7.3)$
        & $9.1\pm 0.3$ $(8.7)$
        & $\,\,\,\,[3\pm 37]$
        & $24^{+4}_{-30}$ $(0)$
        & $16^{+2}_{-21}$ $(0)$ \\
$\overline K {}^0 \eta_8$
        & [$< 4.6$ (90\% CL)]
        & $3.4^{+0.8}_{-0.6}$ $(4.1)$
        & $3.9^{+1.0}_{-0.8}$ $(4.6)$
        & ---
        & $24^{+9}_{-4}$ $(0)$
        & $15^{+3}_{-2}$ $(0)$\\
\hline
$\overline K {}^0\pi^-$
        & $20.6\pm1.3$
        & $19.6^{+2.2}_{-1.4}$ $(18.5)$
        & $21.6\pm 0.6$ $(20.9)$
        & $1\pm 6$
        & $8^{+2}_{-1}$ $(0)$
        & $5\pm 0$ $(0)$\\
$K^-\pi^0$
        & $12.8\pm1.1$
        & $11.6^{+0.5}_{-1.0}$ $(12.1)$
        & $11.0\pm 0.3$ $(10.9)$
        & $1\pm 12$%\footnotemark[1]
        & $-19^{+4}_{-7}$ $(7)$
        & $-14^{+1}_{-2}$ $(6)$\\
$K^-\eta_8$
        & $[3.2\pm0.7]$
        & $3.6^{+0.8}_{-0.7}$ $(4.2)$
        & $4.6^{+1.0}_{-0.8}$ $(5.4)$
        & $[-32\pm20]$
        & $\,\,\,\, 33^{+15}_{-9}$ $(-9)$
        & $\,\,\,\, 19^{+6}_{-4}$ $(-5)$\\
\hline
$\pi^-\pi^0$
        & $5.3\pm0.6$
        & $4.4^{+1.2}_{-0.6}$ $(4.4)$
        & $3.2^{+0.1}_{-0.2}$ $(3.2)$
        & $\,\,\,\,[-7\pm 14]$
        & $0$ $(0)$
        & $0$ $(0)$ \\
\hline
$\pi^-\eta_8$
        & $[3.9\pm0.8]$
        & $1.2^{+0.1}_{-0.3}$ $(1.4)$
        & $1.5^{+0.0}_{-0.1}$ $(1.8)$
        & $[-51\pm19]$
        & $\,\,\,\, 75^{+25}_{-18}$ $(-32)$
        & $\,\,\,\, 42^{+9}_{-6}$ $(-19)$\\
$K^-K^0$
        & $< 2.2$ (90\% CL)
        & $1.7^{+0.3}_{-0.2}$ $(1.5)$
        & $1.3\pm 0.1$ $(1.0)$
        & ---
        & $-84^{+9}_{-14}$ $(-4)$
        & $-79\pm3$ $(-3)$\\
\hline
$\pi^- \pi^+$
        & $4.5\pm0.4$
        & $4.7^{+0.7}_{-0.8}$ $(7.4)$
        & $5.1^{+0.3}_{-0.4}$ $(8.7)$
        & $[\,\,\,\,51\pm 23]$%\footnotemark[2]
        & $A_{\pi\pi}=12^{+14}_{-50}$ $(-24)$
        & $\,\,\,\,13\pm 2$ $(-17)$\\
        &
        &
        &
        & $[-49\pm 61]$%\footnotemark[2]
        & $S_{\pi\pi}= -15^{+69}_{-0}$ $(-5)$
        & $-90^{+1}_{-0}$ $(-88)$\\
$\pi^0\pi^0$
        & $1.7\pm 0.6$
        & $2.5^{+0.4}_{-0.9}$ $(0.2)$
        & $2.8^{+0.3}_{-0.7}$ $(0.1)$
        & ---
        & $-56^{+1}_{-16}$ $(1)$
        & $-35^{+1}_{-6}$ $(1)$\\
$K^-K^+$
        & $<0.6$ (90\% CL)
        & $0.2^{+0.4}_{-0.2}$ $(0.0)$
        & $0.6^{+0}_{-0.1}$ $(0.0)$
        & ---
        & $-13^{+9}_{-76}$ ( --- )
        & $-11^{+2}_{-1}$ ( --- )\\
$\overline K {}^0K^0$
        & $<1.6$ (90\% CL)
        & $1.5^{+0.3}_{-0.6}$ $(1.5)$
        & $1.1\pm 0.1$ $(1.0)$
        & ---
        & $-63^{+141}_{-24}$ $(-4)$
        & $-86^{+6}_{-1}$ $(-3)$\\
$\pi^0\eta_8$
        & ---
        & $0.4^{+0.2}_{-0}$ $(0.3)$
        & $0.2\pm0.0$ $(0.2)$
        & ---
        & $-3\pm0$ $(-4)$
        & $-3\pm0$ $(-4)$\\
$\eta_8\eta_8$
        & ---
        & $0.2\pm0.0$ $(0.1)$
        & $0.2\pm0.1$ $(0.1)$
        & ---
        & $-10^{+96}_{-75}$ $(-6)$
        & $-91^{+14}_{-3}$ $(-6)$\\
\noalign{\smallskip}\hline
\end{tabular}
\end{center}
%\end{ruledtabular}
\end{table*}

\begin{table}[b]
\caption{ The $\chi^2_{\rm min.}$ and fitted parameters for Fits
1, 2. We  constrain $f_\pi F_0^{BK}/f_K F_0^{B\pi} = 0.9\pm 0.1$,
and $1/m_s^{\rm eff}$ gives the effective chiral enhancement for
$\langle O_6\rangle$. The last column is for $\phi_3$ free (fixed)
without FSI phases.
 \label{tab:phase} }
%\begin{ruledtabular}
\begin{center}
\begin{tabular}{crrc}
\hline\noalign{\smallskip}
      &Fit 1
      &Fit 2
      &No FSI
      \\
\hline\noalign{\smallskip}
 $\chi^2_{\rm min.}/{\rm d.o.f.}$
        & $\,\,17/5$
        & $\,\,25/6$
        &$50/7$ ($65/8$)
       \\
 $\phi_3$
        & \ $(96^{+21}_{-10})^\circ$
        & \ $[\,60^\circ]$
        &$106^\circ$ ($[\,60^\circ]$)
       \\
 $\delta$
        &  $(67^{+21}_{-11})^\circ$
        &  $(63^{+6}_{-0})^\circ$
        & ---
        \\
 $\sigma$
        & $(90^{+14}_{-59})^\circ$
        &$(103^{+0}_{-4})^\circ$
        & ---
        \\
 $F_0^{B\pi}$
        & $0.29^{+0.04}_{-0.02}$
        & $0.24^{+0.00}_{-0.01}$
        &0.25 (0.16)
    \\
 $F_0^{BK}$
        & $0.33^{+0.05}_{-0.06}$
        & $0.27^{+0.00}_{-0.02}$
        &0.27 (0.18)
    \\
 $m_s^{\rm eff}$ (MeV)
        & $81^{+26}_{-14}$
        &  $57^{+1}_{-3}$
        & 66 (35)
    \\
\noalign{\smallskip}\hline
\end{tabular}
%\end{ruledtabular}
\end{center}
\end{table}

\section{Results} \label{result}

In Table~\ref{tab:output}, the numbers given in parentheses are by
setting $\delta$ and $\sigma$ to zero but keeping all other
parameters as determined by the fit, which indicates the amount of
FSI cross-feed.
The $\chi^2_{\rm min.}$ and fitted parameters are given in
Table~\ref{tab:phase}.
For illustration, we obtain output errors for both Tables by
scanning the $\chi^2\leq \chi^2_{\rm min.}+1$ parameter space.
The fitted rates and CP asymmetries, as well as the fitted
parameters, when compared with data or with theory, gives one a
sense of reasonableness. For example, the $\chi^2_{\rm min.}/{\rm
d.o.f.}$ for Fit 1 and 2 are 17/5 (giving $\phi_3 \cong 96^\circ$)
vs. 25/6, and the former seems better. This can also be seen from
the low $F_0^{BK(\pi)}$ and $m_s^{\rm eff}$ fitted parameters from
Fit~2, when compared with theory. Both fits are much worse without
FSI: $\chi^2_{\rm min.}/{\rm d.o.f.}$ is 50/7 (65/8) for Fit~1
(2), as seen in the last column of Table~\ref{tab:phase}.

%\section{Driving Force for $\delta$ and $\sigma$} \label{drive}

From Table~\ref{tab:phase} we see that, whether one keeps $\phi_3$
fixed or free, the fitted FSI phases $\delta$ and $\sigma$ are
rather sizable, while disallowing them gives much poorer $\chi^2$.
Let's see what drives these phases.

%Don't forget to give each section and subsection a unique label
%(see Sect.~\ref{sec:1}).
%

\begin{figure}[b]
\resizebox{0.50\textwidth}{!}{%
  \includegraphics{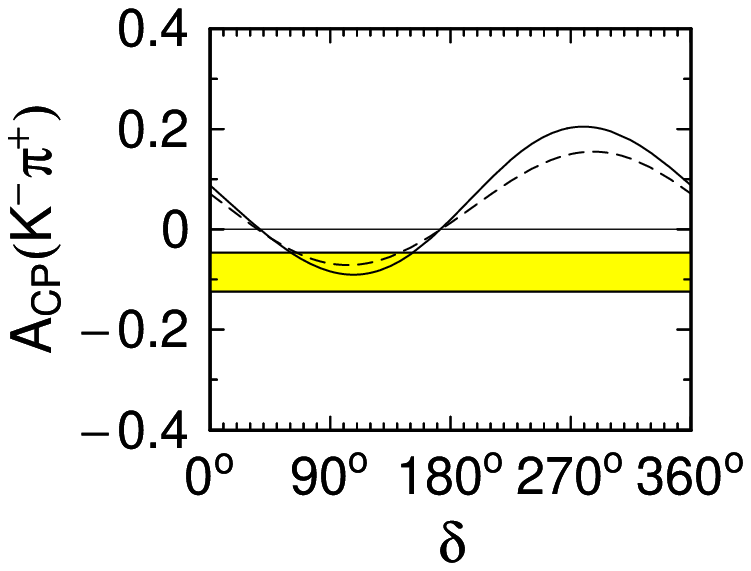} \hskip-0.5cm
  \includegraphics{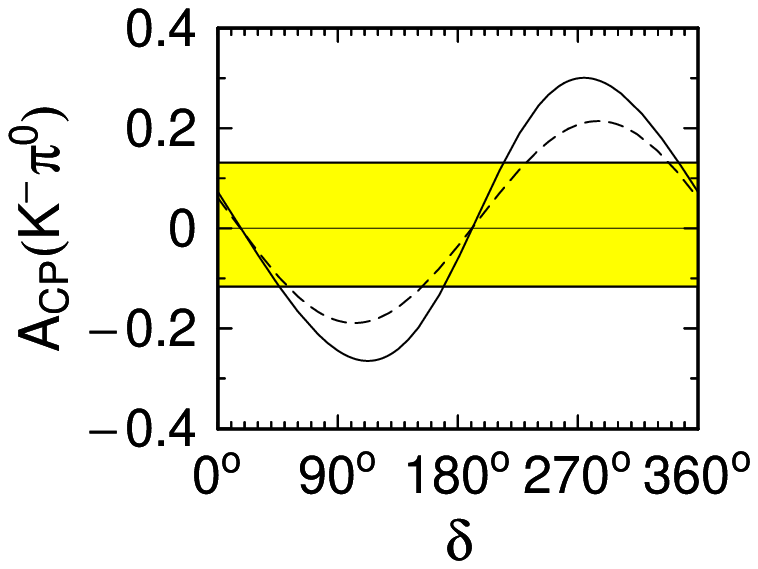}
}
%\vspace{5cm}       % Give the correct figure height in cm
\caption{$A_{\rm CP}(K^-\pi^{+,0})$ vs. $\delta$. Solid (dashed)
line is for Fit~1 (Fit~2), and shaded bands are 1$\sigma$
experimental ranges.}
\label{fig:acp}       % Give a unique label
\end{figure}

%%
%% For two-column wide figures use
%\begin{figure*}
%% Use the relevant command for your figure-insertion program
%% to insert the figure file. See example above.
%% If not, use
%\vspace*{5cm}       % Give the correct figure height in cm
%\end{figure*}
%%

The $\delta$ dependence of $A_{\rm CP}$s for $K^-\pi^+$ and
$K^-\pi^0$ are plotted in Fig.~\ref{fig:acp}. The former now has
some significance, but {\it opposite in sign} w.r.t. QCD
factorization predictions. We see that FSI can bring about a sign
change, which disfavors $\sin\delta < 0$. Together with
``restraint" from $K^-\pi^0$ mode, $\delta \sim 60^\circ$ is more
or less settled between the two. Note that $A_{\rm CP}(K^-\pi^0) <
-14\%$ from both fits. This is in contrast with $(1\pm12)\%$ from
current data, which averages out a sizable {\it positive} value of
$(23\pm 11^{+1}_{-4})\%$ from Belle against {\it negative} central
values reported by both BaBar and CLEO. From a theory standpoint,
$A_{\rm CP}(K^-\pi^0)$ should basically track $A_{\rm
CP}(K^-\pi^+)$, as can be seen from fitted output, which should be
tested with more data.

\begin{figure}
\resizebox{0.5\textwidth}{!}{%
\hskip0.6cm
  \includegraphics{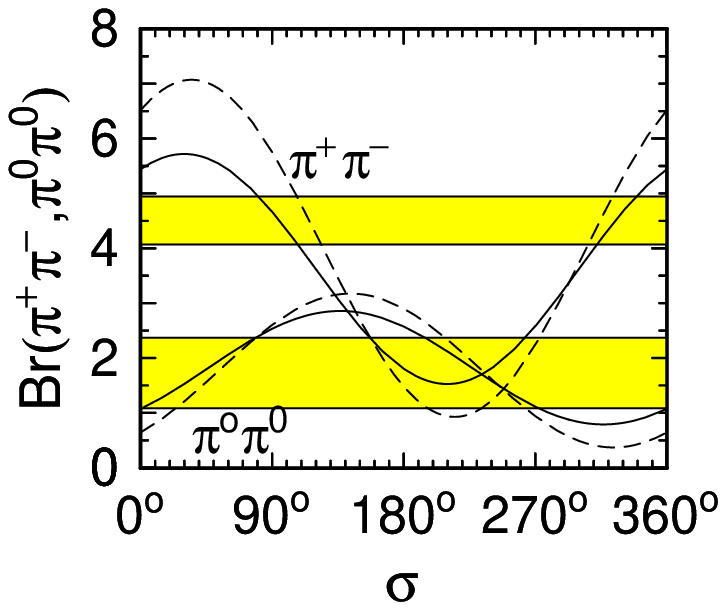} \hskip-0.6cm
  \includegraphics{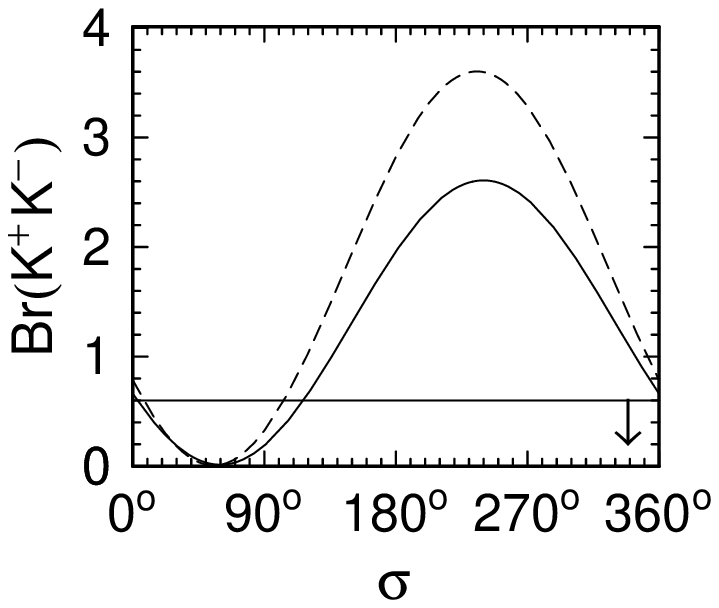}
}
%\vspace{5cm}       % Give the correct figure height in cm
\caption{ Rates ($\times 10^{6}$) for (a) $\pi^+\pi^-$,
$\pi^0\pi^0$, (b) $K^-K^+$ vs. $\sigma$. Solid (dashed) line is
for Fit~1 (Fit~2) with $\delta$ fixed at $67^\circ$ ($63^\circ$).
Horizontal bands are 1$\sigma$ experimental ranges or upper
limit.}
\label{fig:pipikk}       % Give a unique label
\end{figure}

The $\pi^-\pi^+$ and $\pi^0\pi^0$ rates are sensitive to both
$\delta$ and $\sigma$. They are plotted in
Fig.~\ref{fig:pipikk}(a) with $\delta$ fixed to fit values of
Table~\ref{tab:phase}. Both fits clearly favor large $\sigma$, to
account for the smallness of $\pi^-\pi^+$ rate by rescattering
into $\pi^0\pi^0$, which {\it both~\cite{pi0pi0} BaBar and Belle
now have evidence for!}

The rate of $K^-K^+ < 6\times 10^{-7}$ is very suppressed, which
could challenge FSI. From Eq. (\ref{eq:Smatrix}), if $\delta_{\bf
1}$, $\delta_{\bf 8}$, $\delta_{\bf 27}$ are all randomly sizable,
$K^-K^+ \sim \pi^-\pi^+ > 10^{-6}$ would be expected.
However, as seen in Fig.~\ref{fig:pipikk}(b), for $\vert \delta -
\sigma \vert \ltap 50^\circ$, the $K^-K^+$ rate can be comfortably
below the present limit, {\it but $\delta$, $\sigma$ can be
separately large}. The reason is due to the smallness of ${\bf
27}$ in the $I = 0$ $\pi\pi\to \pi\pi$ amplitude.
$K^0\overline K{}^0 \sim K^0K^-\sim 10^{-6}$ are, however, little
perturbed.

Aside from the recent evidence for $\pi^0\pi^0$, the focus of our
interest is the mixing-dependent CP violation in the $\pi^+\pi^-$
mode, where Belle and BaBar have been reporting conflicting
results since 2002. It should be noted that BaBar's 113 fb$^{-1}$
update has moved closer to Belle's, which has yet to update but
published results lie outside the physical region. Both
experiments now agree in the signs, and the new summer 2003
averages are $S_{\pi\pi} = -0.58\pm0.20$ and $A_{\pi\pi} = 0.38\pm
0.16$. In any case, we have not used these numbers in our fit.

\begin{figure}%[b]
\resizebox{0.5\textwidth}{!}{%
  \includegraphics{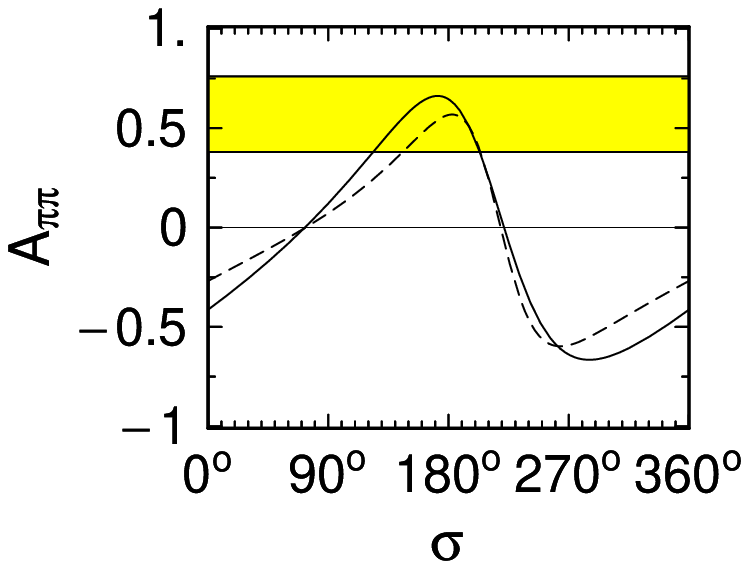} \hskip-0.8cm
  \includegraphics{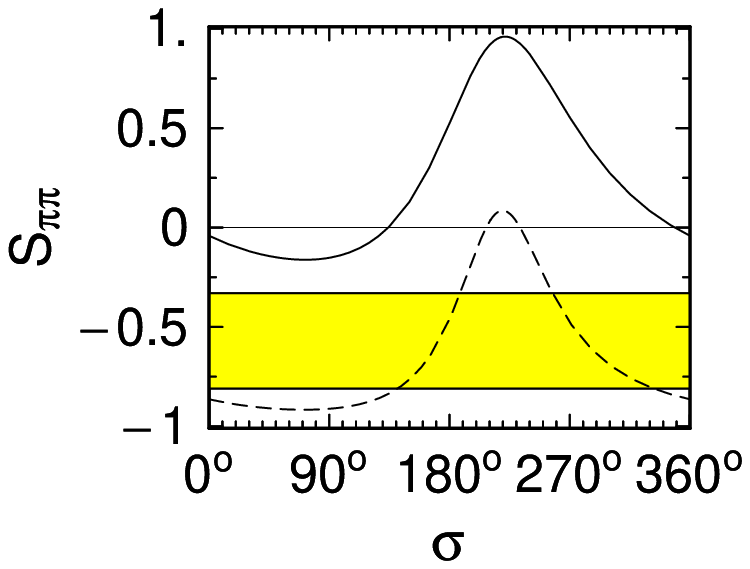}
}
%\vspace{5cm}       % Give the correct figure height in cm
\caption{(a) $A_{\pi\pi}$ and (b) $S_{\pi\pi}$ vs. $\sigma$}
\label{fig:apipispipi}       % Give a unique label
\end{figure}

The $\sigma$ dependence for $A_{\pi\pi}$ and $S_{\pi\pi}$ are
plotted in Fig.~\ref{fig:apipispipi} with $\delta$ fixed to fit
values of Table~\ref{tab:phase}.
As $A_{\pi\pi}$ is nothing but $A_{\rm CP}(\pi^+\pi^-)$, it has
strong FSI dependence. Much like the case of $K^-\pi^+$, large
$\sigma \sim 100^\circ$ turns $A_{\pi\pi}$ positive, of order
10\%, in agreement with BaBar's update number
$-0.19\pm0.19\pm0.05$.
$S_{\pi\pi}$ depends very weakly on $\sigma$ for $\sin\sigma > 0$,
the domain of interest. We have further checked that it is
basically flat in the first quadrant of $\delta$--$\sigma$ plane.
But, as a measure of indirect CP violation, it depends very
strongly on the CP phase $\phi_3$. From
Fig.~\ref{fig:apipispipi}(b) we see that it changes from $\sim 0$
for $\phi_3 \sim 96^\circ$ (Fit 1), and turns almost maximal
negative ($-90\%$) for $\phi_3$ fixed to $60^\circ$ (Fit 2). The
situation is clearly as volatile as the actual competition between
Belle and BaBar!

\section{Discussion} \label{disc}

The situation for $A_{\pi\pi}$ and $S_{\pi\pi}$ is indeed
volatile, as evidenced by the recent shift of BaBar numbers closer
to Belle's, while Belle has yet to update with full 2003 dataset.
It does seem that large FSI is called for, in the form of both
$\delta$ and $\sigma$ being sizable, but $\vert \sigma -
\delta\vert \ltap 50^\circ$ to suppress $K^-K^+$ mode.

The difference between Fits 1 and 2 are marginal, except for
$S_{\pi\pi}$ becoming sizably negative as one moves from
$\phi_3\sim 90^\circ$, towards CKM fit result of $60^\circ$.
However, the latter gives too small a $\pi^-\pi^0$ rate (by
3$\sigma$) and somewhat unreasonably small form factor values.

It is remarkable that in the last two years, the discovery of
sizable color suppressed $D^0h^0$ modes, the appearance of $A_{\rm
CP}(K^-\pi^+) < 0$, evidence for $\pi^0\pi^0$ rate $> 10^{-6}$,
and the emerging (but unfortunately unsettled) picture for
$A_{\pi\pi}$ and $S_{\pi\pi}$, all suggest large FSI together with
large $\phi_3$ may be realized. We find $\delta \sim 60^\circ$ and
$\sigma \sim 100^\circ$ on top of $\phi_3\sim 90^\circ$. To test
these, one needs refined measurement of $\pi^0\pi^0$ rate, as well
as finding $K^-\bar K^0 > 10^{-6}$ but $K^-K^+$ very suppressed.
The value of $A_{\rm CP}(K^-\pi^0) < 0$ should be tested, as well
as $A_{\pi\pi} > 0$. The $S_{\pi\pi}$ parameter would be a good
test for $\phi_3$.

If the two SU(3) rescattering phases differences $\delta$ and
$\sigma$ bear out in the future, it would be a challenge to strong
interaction physics to understand the origin of such large strong
phases. In addition, large $\sigma$ phase implies sizable ``double
annihilation" of initial state flavor content, which is a further
mystery. We have only taken the utilitarian approach of putting
these phases in as parameters. It is amusing that strong
resistance to such strong phases come from not only the QCD
factorization camp, but also from Regge approach camp. Neither
seem willing to admit FSI phases larger than $20^\circ$ or so.

%\begin{table}
%\begin{tabular}{lll}
%\hline
%\end{tabular}
%% Or use
%\vspace*{5cm}  % with the correct table height
%\end{table}

%
% BibTeX users please use
% \bibliographystyle{}
% \bibliography{}

\begin{thebibliography}{}
\bibitem{HHY}
X.G.~He, W.S.~Hou and K.C.~Yang, Phys.\ Rev.\ Lett.\  {\bf 83},
(1999) 1100.
%
\bibitem{HY}
W.S.~Hou and K.C.~Yang, Phys.\ Rev.\ Lett.\  {\bf 84}, (2000)
4806.
%
\bibitem{D0h0}
K.~Abe {\it et al.}  [BELLE Collab.], Phys.\ Rev.\ Lett.\  {\bf
88} (2002) 052002; T.E.~Coan {\it et al.}  [CLEO Collab.], {\it
ibid.}\ {\bf 88} (2002) 062001.
%
\bibitem{LP03}
For latest reults, see the plenary talks by T. Browder, J. Fry,
and H. Jawahery at Lepton Photon Symposium, August 2003, Fermilab,
USA.
%
\bibitem{CHY}
C.K.~Chua, W.S.~Hou and K.C.~Yang, Phys.\ Rev.\ D {\bf 65}, (2002)
096007.
%
\bibitem{CHY2}
C.K.~Chua, W.S.~Hou and K.C.~Yang, Mod.\ Phys.\ Lett.\ A~{\bf 18},
(2003) 1763.
%
\bibitem{HSW}
W.S.~Hou, J.G.~Smith and F.~W\" urthwein, hep-ex/9910014.
%
\bibitem{pi0pi0}
B.~Aubert {\it et al.}  [BABAR Collab.], hep-ex/0308012; K.~Abe
{\it et al.} [Belle Collab.], hep-ex/0308040.

%\bibitem{RefB} Author, \textit{Book title} (Publisher, place
%year) page numbers
% etc

\end{thebibliography}

\end{document}